\documentclass[12pt,prd,amsfonts,nofootinbib,showpacs]{revtex4}
\usepackage{graphicx, epsfig}
\newcommand{\be}{\begin{equation}}
\newcommand{\ee}{\end{equation}}
\newcommand{\ba}{\begin{eqnarray}}
\newcommand{\ea}{\end{eqnarray}}
\newcommand{\nn}{\nonumber}
\newcommand{\x}{\times}
\begin{document}
                                                                                
\bigskip
\vspace{2cm}
\title{$(\omega,\ \phi) P^-$ decays of tau leptons}
\vskip 6ex
\author{A. Flores-Tlalpa}
\email{afflores@fis.cinvestav.mx}
\author{G. L\'opez Castro}
\email{glopez@fis.cinvestav.mx}
\affiliation{Departamento de F\'{\i}sica, Cinvestav,
Apartado Postal 14-740, 07000 M\'exico, D.F., M\'exico}
\bigskip
\begin{abstract}
   \hspace*{5mm} 
The $\tau^- \to (\omega, \phi)P^-\nu_{\tau}$ decays, where $P^-=\pi^-, 
K^-$, are considered within a phenomenological model with dominance of 
meson intermediate states. We assume SU(3) flavor symmetry to fix 
some of the unknown strong interaction couplings. Our predictions for the 
$\tau^- \to \phi (\pi^-,\ K^-)\nu_{\tau}$ branching fractions are in 
good agreement with recent measurements of the BABAR  and BELLE  
Collaborations.
 \end{abstract}


\maketitle

\section{ Introduction}

Tau lepton decays into a charged pseudoscalar $P^-$ and an isoscalar 
vector meson $V$, generically denoted by $\tau^- \to  VP^-\nu_{\tau}$,  
can occur in four possible ways:
\ba
\tau^- &\to & \omega \pi^- \nu\ ,  \\
\tau^- &\to & \phi \pi^- \nu\ , \\
\tau^- &\to & \omega K^- \nu\ ,  \\
\tau^- &\to & \phi K^- \nu\ .  
 \ea
Owing to the quark mixing angle factors, one naively expects that 
processes  (1) and (2) ($\Delta S=0$)  would have larger branching 
fractions than decay modes involving a $K^-$ meson ($\Delta S=-1$). 
However, the rich resonance structure of intermediate states combined with 
the  high thresholds for the above processes will produce an interesting 
pattern worth to be investigated.

 The study of $\tau^- \to VP^-\nu_{\tau}$ decays is interesting for 
several reasons. As is well  known, tau decays into several pseudoscalar 
mesons are dominated by the production of  intermediate resonant 
states \cite{Yao:2006px}. A good quantitatively description of the 
decay modes shown in Eqs. (1)-(4) is  important to better  understand the 
dynamics of three and four pseudoscalar mesons produced in tau lepton 
decays.  On the other hand, the 
study of such  decays allows a direct access  to  the  $\langle 
VP|J_{\mu}|0 \rangle$ hadronic matrix element in the  intermediate energy 
regime. Since $\tau \to VP\nu$ and $B,D \to Vl\nu$ decays are 
related by crossing, they can be useful  to provide further 
tests of either non-relativistic \cite{Isgur:1988gb} or  relativistic 
\cite{Wirbel:1985ji} 
quark model predictions. Finally, the $\tau^- \to (\omega,\phi)\pi^-\nu$ 
decays are related to the $e^+e^- \to  
(\omega,\phi)\pi^0$ processes via isospin symmetry and their measurements 
can be useful to provide another test of the conserved vector current (CVC) 
hypothesis \cite{Decker,Eidelman,LopezCastro:1996xh}.

In Table I we display the experimental values for the branching ratios of 
$\tau^- \to VP^-\nu$ decays. The $\omega\pi^-$ final state is the most 
favored and its  branching  fraction and  spectral function were the first 
to be measured \cite{Yao:2006px, Omega, Edwards:1999fj}. 
\begin{table}[t]
%\begin{table}
\begin{tabular}{ccc}
\hline
$VP^-$ mode & \ \ \ Branching fraction &  Reference \\
\hline \hline 
$\omega \pi^-$ & $(1.95\pm 0.08)\times 10^{-2}$ & 
\cite{Omega,Edwards:1999fj} \\
\hline
$\phi \pi^-$ & $(6.05\pm 0.71)\times 10^{-5}$ & \cite{Abe:2006uk},\\   
& $(3.42\pm 0.55 \pm 0.25)\times 10^{-5}$ & \cite{Babar} \\
\hline 
$\omega K^-$ & $(4.1\pm0.9)\times 10^{-4}$ & \cite{omegak} \\
\hline 
$\phi K^-$ & $(4.05 \pm 0.25\pm  0.26) \times 10^{-5}$ & \cite{Abe:2006uk} 
\\
& $(3.39\pm 0.20\pm 0.28)\times 10^{-5}$ & \cite{Babar} \\
\hline
\end{tabular}
\caption{Measured branching fractions of $\tau^-\to VP^-\nu_{\tau}$ 
decays.} 
\end{table}
Because of their smaller branching fractions, the decay 
modes in  Eqs. (2)-(4) were measured only very recently 
\cite{omegak,Babar,Abe:2006uk}. Previous upper bounds on $\tau$ 
decays involving $\phi$ mesons were reported in Ref. \cite{Avery}, where  
the upper limits  $B(\tau \to \phi \pi \nu ) \leq (1.2\sim 2.0) \times 
10^{-4}$ and 
$B(\tau \to \phi K\nu) \leq (5.4 \sim 6.7) \times 10^{-4}$ were set at 
the 90\% c.l. \cite{Avery}. 

 Earlier theoretical estimates for some of these decays were considered  
in  references \cite{Decker,Eidelman,LopezCastro:1996xh}. Based on 
the  Conserved Vector Current (CVC) hypothesis and using bounds for  
the  cross  section of $e^+e^- \to \phi\pi^0$, the loosely limit $B(\tau^- 
\to  \phi\pi^-\nu)  \leq 
9.0 \times 10^{-4}$ at 90\% c.l. was derived in \cite{Eidelman}. On 
another hand, by assuming that the form factor of this decay is dominated 
by the contribution of two vector resonances ($\rho$ and $\rho'$) and 
that flavor SU(3) is a good symmetry, the value $B(\tau^- \to 
\phi\pi^-\nu) 
= (1.20\pm 0.48) \times 10^{-5}$ was obtained in 
Ref. \cite{LopezCastro:1996xh}. 
This prediction clearly underestimates the measured fraction of 
$\phi\pi^-$ (see Table I). Concerning the $\phi K^-$ modes only rough 
estimates are available based on phase-space and quark mixing angles 
considerations \cite{Abe:2006uk}. In this paper we revisit this 
subject and provide a unified  description of the four  $VP^-$ decays 
shown in Eqs. (1)-(4) in the framework of a meson dominance  model. 

In this paper we consider the possibility that a meson dominance model 
with a few  intermediate states can account, in a unified way, for the 
observed branching  fractions of reactions (1)-(4). As a simplifying 
assumption we will rely on SU(3) flavor symmetry for the strong couplings 
and we will assume an ideal value $\tan \theta_V=1/\sqrt{2}$ 
of the $\omega-\phi$ mixing angle ({\it i.e.}, $\phi$ is assumed to be an  
almost pure $\bar{s}s$ state). On the basis of these assumptions we 
conclude that present data on $\tau^- \to 
VP^-\nu$ decays can be easily  accommodated within the meson dominance 
model.

\section{ Meson dominance model for tau decays}  
   Thus, let us first consider the decay $\tau^- (p_{\tau}) \to 
V(p_V)P^-(p_P)\nu (p_{\nu})$, where $p_i$ denote the four-momenta of 
particle $i$. The hadronic matrix element can be decomposed in terms of 
four form factors ($Q=s$ or $d$) \cite{Scora:1989ys}:
\be
\langle VP|\bar{Q}\gamma^{\alpha}(1-\gamma_5)u|0 \rangle =
ig\varepsilon^{\alpha\beta \mu\nu} \epsilon^*_{\beta}q_{+\mu}q_{-\nu} 
 + f\epsilon^{*\alpha} 
+[a_+q_+^{\alpha}+a_-q_-^{\alpha}]\epsilon^*\cdot q_+\, 
\ee
where $\epsilon^*_{\beta}$ is the polarization four-vector of the outgoing 
vector meson ($p_V\cdot \epsilon^*=0$), and $q_{\pm} = p_V\pm p_P$. The 
vector  ($g$) and axial ($f,\ 
a_{\pm}$) form factors are functions of $s=q_+^2$ only. 

 If we define $\Sigma^2=m_V^2+m_P^2,\ \Delta^2=m_V^2-m_P^2$,
and $\beta_{VP}=(1-2\Sigma^2/s+\Delta^4/s^2)^{1/2}$,  the 
differential decay rate can be written in the simple form:
\ba
\frac{d\Gamma}{ds} &=& \frac{G_F^2|V_{uQ}|^2m_{\tau}^3}{128\pi^3} 
\beta_{VP} \left(1-\frac{s}{m_{\tau}^2}\right)^2 \nonumber \\
&& \times \left\{ \frac{1}{2}\beta_{++} + \frac{1}{2}\beta_{--}\left[ 
\frac{\Delta^4}{s^2} + \frac{1}{3} \left(1+\frac{2s}{m_{\tau}^2}\right) 
\beta_{VP}^2 \right] + \frac{\Delta^2}{s}Re[\beta_{+-}] + 
\frac{\alpha}{m_{\tau}^2} \right\} \ , 
\ea
where $G_F$ is the Fermi constant,  $V_{uQ}$ is the $uQ$ entry of the 
Cabibbo-Kobayashi-Maskawa matrix  and, 
\ba 
\beta_{++}&=&\frac{1}{4m_{V}^2} 
\left\{|f|^2+4m_V^2|g|^2(s-2\Sigma^2)+|a_+|^2s^2\beta_{VP}^2 + 
2Re[fa_+^*]2(s-3m_V^2-m_P^2) \right\} \nonumber \\
\beta_{--}&=&\frac{1}{4m_{V}^2} 
\left\{|f|^2-4m_V^2s|g|^2+|a_-|^2s^2\beta_{VP}^2+2Re[fa_-^*](s+\Delta^2) 
\right\} \nonumber \\
\beta_{+-}&=& \frac{1}{4m_V^2} \left \{ 
|f|^2+4m_V^2\Delta^2|g|^2+(fa_+^*)^*(s+\Delta^2)+(fa_-^*) 
(s-3m_V^2-m_P^2) \right. \nonumber \\
&& \left. \  \ \ \ \ \ \ \ \ \ +(a_+a_-^*)s^2\beta_{VP}^2 \right\} 
\nonumber \\
\alpha &=& |f|^2+s^2|g|^2\beta_{VP}^2 \ . 
\ea

The Feynman diagrams with the intermediate mesons that connect the 
weak current and the strong vertex in $\tau \to VP\nu$ decays are shown in 
Figure 1. 
\begin{figure}
\includegraphics[width=12cm]{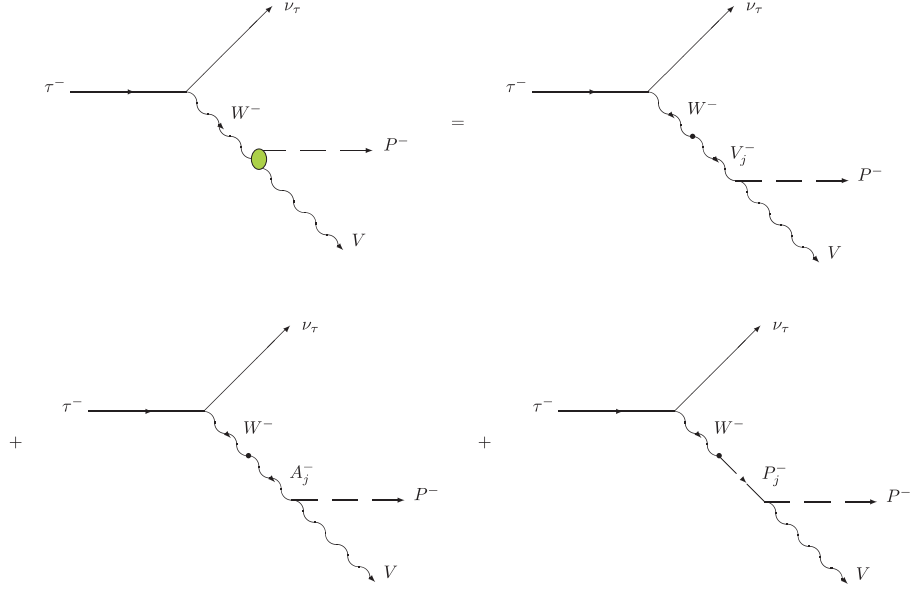}
%\vspace{-10.0cm}
\caption{Intermediate virtual meson contributions to $\tau^- \to 
VP^-\nu$ decays}
\end{figure}
Using the Feynman rules for the elements of these diagrams, we 
get the following expressions for the form factors ($V_j,\ A_j$ and $P_j$ 
denote vector, axial and pseudoscalar intermediate meson states, 
respectively):
\ba
g&=& \frac{1}{2} \sum_j \frac{f_{V_j}g_{V_jVP}}{D_{V_j}(s)}\ , \nonumber 
\\
f&=& -\frac{1}{2}(s+\Delta^2)\sum_j\frac{f_{A_j}g_{A_jVP}}{D_{A_j}(s_)}\ ,  
\nonumber \\
a_+&=& \frac{1}{2}\sum_j\frac{f_{A_j}g_{A_jVP}}{D_{A_j}(s_)} + 2\sum_j 
\frac{f_{P_j}g_{P_jVP}}{D_{P_j}(s)}\ , \nonumber \\
a_-&=& \frac{1}{2}\sum_j \frac{f_{P_j}g_{P_jVP}}{D_{P_j}(s)}\ ,
\ea
where $f_{M_j}$ denotes the weak coupling of the  $M_j$ intermediate 
meson, $g_{M_jVP}$ is its strong coupling to the $VP$ final 
state, and $D_{M_j}(s) \equiv s-m_{M_j}^2+im_{M_j}\Gamma_{M_j}$, where 
$m_{M_j}\ (\Gamma_{M_J})$ is the mass (width) parameter of the 
corresponding intermediate state.

\section{Strangeness-conserving decays}

The G-parity properties of the weak currents and $V\pi^-$ system 
in  this case, impose $f=a_-=a_+=0$. As in previous papers 
\cite{Decker,LopezCastro:1996xh}, we will 
assume that the non-vanishing vector form factor is saturated by the 
exchange of two vector resonances (the $\rho(770)$ and the 
$\rho'(1523)$). Thus we get ($V=\omega,\ \phi$):
\be
g(s)=\frac{f_{\rho}g_{\rho V \pi}}{2D_{\rho}(s)} 
\left\{1+\alpha_{V\pi} \frac{D_{\rho}(s)}{D_{\rho'}(s)} \right\}\ ,
\ee
where $\alpha_{V\pi}=f_{\rho'}g_{\rho' V \pi}/f_{\rho}g_{\rho V \pi}$ is 
the only free parameter of the model at this stage. We chose the 
$\rho'(1523)$ state ($m_{\rho'}=1523$ MeV and $\Gamma_{\rho'}=400$ MeV 
\cite{Edwards:1999fj, Akhmetshin:2003ag, Davier:2005xq}) as the second 
vector 
resonance,  instead of the 
$\rho(1450)$ \cite{Yao:2006px}, because its larger width allows for a 
better  fit to 
data on the spectral function \cite{Edwards:1999fj, 
Akhmetshin:2003ag, Davier:2005xq}.

As usual \cite{Davier:2005xq}, we can define a vector spectral function 
whose expression becomes very simple in this case:
\be
v(s)=\frac{s\beta^3_{V\pi}}{12\pi} \left| g(s)\right|^2\ .
\ee
This spectral function has been  measured by the  ALEPH \cite{Omega} and 
CLEO \cite{Edwards:1999fj} Collaborations  for the dominant $\omega\pi^-$ 
final state. In order to fit the data on the spectral function, we use: 
$f_{\rho}= (170.0 \pm 
3.4)\times 10^3$ MeV$^2$ and $g_{\rho\omega\pi}=(15.2\pm 1.9)\times 
10^{-3}$ MeV$^{-1}$  (this value is a bit larger than the average value
$g_{\rho\omega\pi}=(12.3\pm 1.2)\times 10^{-3}$ MeV$^{-1}$ obtained from 
$\rho^{\pm} \to  \pi^{\pm}\gamma,\ \rho^0\to \pi^0\gamma$ 
and $\omega \to \pi^0\gamma$ decays, but they agree within their error 
bars). A  fit to the 
spectral function reported in Ref. \cite{Omega} gives us   
$\alpha_{\omega \pi}= -0.57 \pm 0.11$. The data of Ref. 
\cite{Edwards:1999fj} and  the fitted curves of the spectral function are 
shown in Figure 2. 

\begin{figure}
\includegraphics[width=11cm]{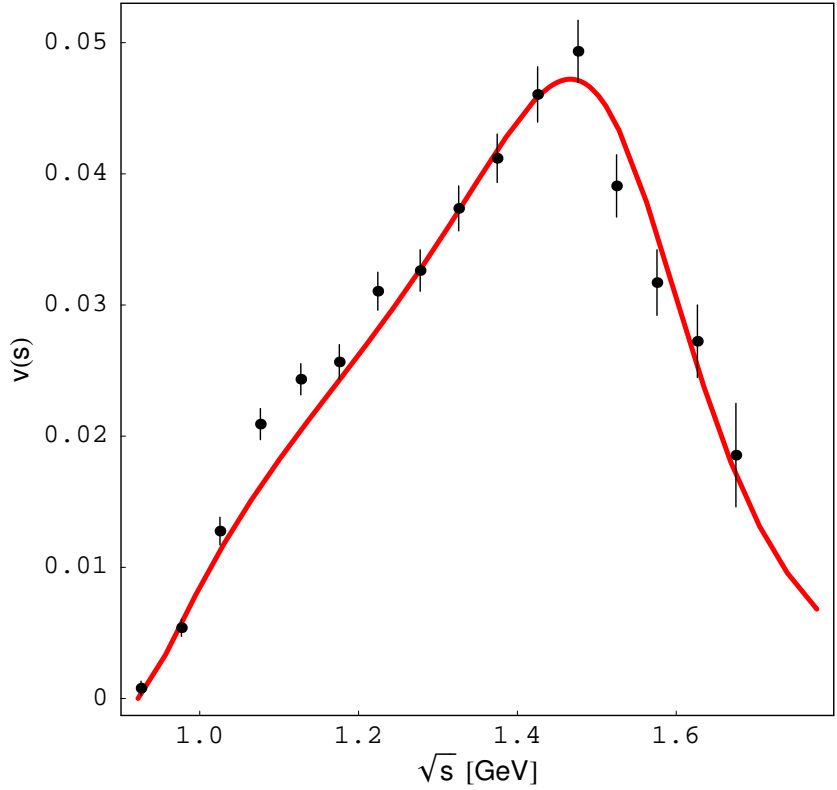}
%\vspace{-10.0cm}
\caption{Spectral function $v(s)$ of the $\omega \pi^-$ system: the 
best fit is represented by the solid line; the experimental data from 
CLEO \cite{Edwards:1999fj} are shown with solid dots. } \end{figure}

Using the above value of $\alpha_{\omega\pi}$ we can derive the following 
branching fraction by integration of Eq. (6):
\be
B(\tau^- \to \omega\pi^-\nu_{\tau}) = (1.95\pm 0.60)\% \ ,
\ee
which is in very good agreement with the experimental value shown in Table 
I.

   Now we focus on the $\phi\pi$ decay channel. We will assume that flavor 
SU(3) is a good symmetry for the $VVP$ couplings of the octet of  vector 
mesons and of their radial excitations. Under this assumption, we can 
get $\alpha_{\phi\pi}=\alpha_{\omega\pi}$ for the relative weights of 
$\rho$ and $\rho'$ contributions in Eq. (9). In addition we use 
$g_{\rho\phi\pi}=-(1.57\pm 0.03)\times 10^{-3}$ MeV$^{-1}$, which is  
obtained from the $\phi \to \rho\pi \to \pi^+\pi^-\pi^0$ decay rate 
\cite{Yao:2006px}. Now, if we insert these parameters into Eq. (6), we 
get:
\be
B(\tau^- \to \phi\pi^-\nu_{\tau}) = (3.64\pm 0.93) \times 10^{-5} \ ,\\
\ee
which clearly favors the result reported by BABAR \cite{Babar} (see Table 
I).

\section{ Coupling constants for $\Delta S=-1$ decays}

The case of $\Delta S=-1$ tau decays is more difficult to deal with  
because the vector and axial weak currents contribute to the decay 
amplitude. 
Consequently, more independent information is needed to specify the input 
coupling constants.  
The vector form  factor $g$ will be assumed to be dominated by the 
$K^*=K^*(892)$ and the 
$K'^*=K^*(1410)$ intermediate vector mesons (the coupling of the 
$K''^*=K^*(1680)$ to the $VK^-$ 
system is more suppressed). The axial form factors $f,\ a_{\pm}$ will 
originate from the exchange of the $K^-$ pseudoscalar and the 
$K_1=K_1(1270)$, $K'_1=K_1(1400)$ axial mesons. The  expressions for the 
form factors were given in Eq. (9), and the numerical values of coupling
constants will be discussed in the following subsections.

\subsection{Weak couplings}
The value of the $K^-$ weak coupling is known with good precision, 
$f_K=(159.8 \pm  1.5)$ MeV \cite{Yao:2006px}, and is extracted from $K \to 
\mu\nu$  decays by including the effects of radiative corrections. The 
weak couplings of the other mesons can be extracted from the measurements  
of $\tau \to  {\cal K}\nu$ decays (where ${\cal K}$ denotes either state 
among $K^*, K'^*, K_1, K'_1$ mesons). We get: 
\ba
f_{K^*}&=& (188.9\pm  4.1)\x 10^3 \ {\rm MeV}^2\ , \nn \\
 f_{K'^*}&=& (170^{+80}_{-57}) \x 10^3\ {\rm MeV}^2\ , \nn \\ 
f_{K_1}&=&(215\pm 25)\x 10^3\ {\rm MeV}^2\ , \nn \\ 
 f_{K'_1}&=& (170 \pm  130)\x 10^3\ {\rm MeV}^2\ . 
\ea
The weak coupling with the largest uncertainty corresponds to the $K'_1$ 
meson. Since the contribution of the $K'_1$ intermediate state will 
provide an important contribution, a more precise measurement of the $\tau 
\to K_1(1400)\nu$ would be suitable to improve the accuracy of our 
prediction.

\subsection{$KVK$ couplings}

The determination of the strong couplings of the intermediate  resonances 
to the $VK^-$ system in a reliable way is also a difficult task, 
either  because such decays are not allowed by kinematics or because 
there are not independent processes where their contribution can be 
studied. Thus we will strongly rely on 
SU(3) flavor symmetry to fix their values when necessary. 

From the experimental value of the $\phi \to K^+K^-$ branching 
fraction \cite{Yao:2006px}, we get $g_{K^+\phi K^-}=(4.48\pm 0.04)$.  
Now, using the SU(3) symmetry and assuming an ideal value for the 
$\omega-\phi$ mixing angle ($\tan\theta_V =1/\sqrt{2}$), we get:
\be 
g_{K^+\omega K^-}= g_{K^+\phi K^-}\tan\theta_V=(3.17 \pm 0.03). 
\ee
An alternative calculation of the $KVK$ couplings can be obtained 
by assuming  the  vector-meson dominance model of the kaon electromagnetic 
form  factors at zero momentum transfer:
\ba
F_{K^+}(0) &=& \frac{g_{K^+\rho^0K^+}}{\gamma_{\rho}} + 
\frac{g_{K^+\omega K^+}}{\gamma_{\omega}}+ \frac{g_{K^+\phi 
K^+}}{\gamma_{\phi}} =1 \\
F_{K^0}(0) &=& \frac{g_{K^0\rho^0K^0}}{\gamma_{\rho}} +
\frac{g_{K^0\omega K^0}}{\gamma_{\omega}}+ \frac{g_{K^0\phi
K^0}}{\gamma_{\phi}} =0\ ,
  \ea
where $em_V/\gamma_V$ defines the coupling of the neutral vector meson $V$ 
to the  photon. Now, we can use the SU(3) relations between the $PVP'$ 
couplings (we assume again $\tan \theta_V=1/\sqrt{2}$):
\ba
g_{K^+ \rho^0 K^-} &=& -g_{K^0\rho^0 \bar{K}^0} = \frac{1}{2} G^8_{PVP'}\ 
, 
\nonumber \\
g_{K^+ \omega K^-} &=& g_{K^0\omega \bar{K}^0} = \frac{1}{\sqrt{2}} 
G^8_{PVP'}\ , \nonumber \\
g_{K^+ \phi K^-} &=& g_{K^0\phi \bar{K}^0} = \frac{1}{2} G^8_{PVP'} .
\ea 
Solving the set of eqs. (14)-(16), we finally get:
\ba
g_{K^+ \omega K^-} &=& 
\frac{\gamma_{\omega}\gamma_{\phi}}{2(\gamma_{\phi}+ 
\sqrt{2}\gamma_{\omega})} = 2.99 \pm 0.13 \nonumber \\
g_{K^+ \phi K^-} &=& 
\frac{\gamma_{\omega}\gamma_{\phi}}{\sqrt{2}(\gamma_{\phi}+ 
\sqrt{2}\gamma_{\omega})} = 4.24 \pm 0.19 \ , 
\ea
which are quite similar values to the ones computed from the $\phi \to KK$ 
decays (see above).

\subsection{$V'VP$ strong couplings}
Flavor SU(3) symmetry  predicts the following relations among $V'VP^-$ 
couplings (we assume later below $\tan \theta_V=1/\sqrt{2}$):
\ba
g_{K^*\omega K}&=& -\frac{1}{2\sqrt{3}} G^8_{VV'P}[\sin 
\theta_V-2r\sqrt{2} \cos \theta_V]\ , \\
g_{K^*\phi K}&=& -\frac{1}{2\sqrt{3}} G^8_{VV'P}[\cos\theta_V+ 
2r\sqrt{2}\sin\theta_V]\ , \\
g_{\rho\omega \pi}&=& \frac{1}{\sqrt{3}} G^8_{VV'P}[\sin 
\theta_V+\sqrt{2}r \cos \theta_V]\ , \\
g_{\rho\phi \pi}&=& \frac{1}{\sqrt{3}} G^8_{VV'P}[\cos 
\theta_V- \sqrt{2}r \sin \theta_V]\ ,
\ea
where $r\equiv G^0_{V'VP}/G^8_{V'VP}$ is the ratio of SU(3)  
singlet and octet $V'VP$ couplings. The value of $r$ can be obtained  
from the ratio  of Eqs. (21,22) using the values of $g_{\rho \omega\pi}$ 
and 
$g_{\rho\phi\pi}$ given in the previous section. In this way we get 
$r=0.1.256 \pm 0.038$. If we insert now this value of $r$ into eqs. 
(19,20) 
and use the experimental value of $g_{\rho\omega\pi}$ (see previous 
section) and the ideal value of the $\omega-\phi$ mixing angle, we get:
\ba
g_{K^*\omega K} &=& \left[\frac{4r-1}{4r+2} \right] g_{\rho\omega\pi} = 
(8.71 \pm 0.95) \times 10^{-3}\ \mbox{\rm MeV}^{-1} \ , \\
g_{K^*\phi K} &=& -\frac{1}{\sqrt{2}}g_{\rho\omega\pi} = 
-(10.7 \pm 1.3) \times 10^{-3}\ \mbox{\rm MeV}^{-1} \ .
\ea

\subsection{$AVP$ couplings }

The couplings of axial-vector mesons ($A$) to the $VK^-$ system are the 
most difficult to determine. One may attempt to compute them from the 
measured branching fractions of $K_1, K'_1$ into the $\omega K^-$ channel 
(which is the  only allowed by kinematics). We get from this\footnote{In 
order to extract $g_{K_1\omega K}$ we have taken the maximum values for 
the mass and  width of $K_1$ that are allowed by their error bars 
\cite{Yao:2006px}.}: 
\ba
g_{K_1\omega K} &=& -(3.17 \pm 0.46) \times 10^{-3} \mbox{\rm MeV}^{-1} \ 
,  \\
g_{K'_1\omega K} &=& (4.8 \pm 2.4) \times 10^{-4} \mbox{\rm MeV}^{-1} \ .
\ea
However, the $K_1\phi K$ couplings can not be obtained in this way given 
that $K_1, K'_1 \to \phi K$ decays are not allowed by kinematics.

Given the poor quality of the measurements used to extract the above 
couplings,  we can resort again to the SU(3) flavor symmetry. Since the 
$(K_1,\ K'_1)$ 
physical states are a mixture of the $(K_{1A},\ K_{1B})$ states that 
belong to different $1^3P_1$ ($A$) and $1^1P_1$ ($B$) multiplets of axial 
mesons, we propose as the starting  point the following ${\cal A}VP$ 
interaction  Lagrangian:
\ba
{\cal L}_{{\cal A} V P} & =  & i g_{A V P}^8 f_{a b c} P^a
(\partial^\alpha A^{b \beta})  (\partial_\alpha V_\beta^c -
\partial_\beta
V_\alpha^c) + g_{B V P}^8 d_{a b c} P^a  (\partial^\alpha B^{b \beta})
(\partial_\alpha V_\beta^c - \partial_\beta  V_\alpha^c) \nn \\ & & +
\sqrt{\frac{2}{3}} g_{B V P}^0 \delta_{a b}  P^a (\partial^\alpha B^{b
\beta}) (\partial_\alpha V_\beta^0 - \partial_\beta V_\alpha^0) \ .
\ea
The physical strange axial mesons are defined in terms of flavor SU(3) 
states as follows:
\ba
K_1 &=& K_{1B}\cos \alpha-K_{1A}\sin \alpha\ , \ \ \  ({\rm for}\  K_1^+,\ 
K_1^0)  \nn \\ 
K'_1 &=& K_{1B}\sin \alpha+K_{1A}\cos \alpha\ , \ \ \ ({\rm for}\  
K^{'+}_1,\ K^{'0}_1)  
\ea
and 
\ba
\bar{K}_1 &=& -\bar{K}_{1B}\cos \alpha-\bar{K}_{1A}\sin \alpha\ , \ \ \  
({\rm  for}\   K_1^-,\ \bar{K}_1^0)  \nn \\
\bar{K}'_1 &=& -\bar{K}_{1B}\sin \alpha+\bar{K}_{1A}\cos \alpha\ , \ \ \ 
({\rm for}\ K^{'-}_1,\ \bar{K}^{'0}_1)  \ . 
\ea

The determination of the $K_{1A}-K_{1B}$ mixing angle is still 
controversial \cite{Cheng:2003sm}. According to different authors 
its value can be in the range $30^0 \leq \alpha \leq 60^0$  
\cite{Cheng:2003sm}. 
For illustrative purposes, in the present paper we will use $\alpha = 
45^0$ \cite{Yao:2006px} . If in addition we assume a nonet symmetry for 
the $1^1P_1$ 
couplings, namely $g_{BVP}^8=g_{BVP}^0$ and the ideal value for the 
$\omega-\phi$ mixing angle ($\tan \theta_V=1/\sqrt{2}$), we get the 
following simplified expressions for the couplings that 
involve the $\phi$ and $\omega$ mesons (the expressions of the couplings 
constants for arbitrary values of the $\alpha$ and $\theta_V$ mixing 
angles are given in the Appendix):
\ba
g_{K_1^+ \omega K^-} & = & g_{K_1^0 \omega \overline{K}^0} = - g_{K_1^-
\omega K^+} = - g_{\overline{K}_1^0 \omega K^0} = \frac{1}{2 \sqrt{2}}
\Sigma_+ \ , \\
g_{K_1^{'+} \omega K^-} & = & g_{K_1^{'0} \omega \overline{K}^0} = -
g_{K_1^{'-} \omega K^+} = - g_{\overline{K}_1^{'0} \omega K^0} = -
\frac{1}{2 \sqrt{2}} \Sigma_- \ , \\
g_{K_1^+ \phi K^-} & = & g_{K_1^0 \phi \overline{K}^0} = - g_{K_1^- \phi
K^+} = - g_{\overline{K}_1^0 \phi K^0} = \frac{1}{2} \Sigma_- \ , \\
g_{K_1^{'+} \phi K^-} & = & g_{K_1^{'0} \phi \overline{K}^0} = -
g_{K_1^{'-} \phi K^+} = - g_{\overline{K}_1^{'0} \phi K^0} = -
\frac{1}{2} \Sigma_+ \ ,
\ea
where we have defined $\Sigma_{\pm} \equiv g^8_{AVP}\pm g^8_{BVP}$.

  We can fix the values of the effective couplings $\Sigma_{\pm}$ by 
using the decay rates of $K_1, K'_1$ axial mesons in the same limit where 
Eqs. (30)-(32) were obtained. Using the expressions given in the Appendix 
and comparing with the measured rates of $K'_1 \to K^*\pi$ decays 
\cite{Yao:2006px}, we obtain: $\Sigma_+ = (5.50 \pm 0.27)\x 10^{-3}$ 
MeV$^{-1}$. Similarly, from the measured rate of $K'_1 \to \omega K$ we 
get:  $\Sigma_- = (1.36 \pm 0.67)\x 10^{-3}$ MeV$^{-1}$. Finally, if we 
insert  these values into Eqs. (30)-(33), we get:
\ba
g_{K_1^- \omega K^-} & = & - (1.94 \pm 0.10)  \times
10^{-3} \, {\rm MeV}^{-1} \ , \\
g_{K_1^{'-} \omega K^-} & = & (4.8 \pm 2.4)  \times
10^{-4} \, {\rm MeV}^{-1} \ , \\
g_{K_1^- \phi K^-} & = & - (6.8 \pm 3.4)  \times 10^{-4}
\, {\rm MeV}^{-1} \ , \\
g_{K_1^{'-} \phi K^-} & = & (2.75 \pm 0.14)  \times
10^{-3} \, {\rm MeV}^{-1} \ .
\ea
Observe that the $K_1\omega K$ coupling in Eqs. 
(34) and (25) have similar sizes despite the different sources 
used for their determination. In our calculations we will use the 
numerical values shown in Eqs. (34)-(37).

\section{ Strangeness-changing decays}

  With the information on the coupling constants given in the previous 
section, the only free parameters to our disposal are the relative 
contributions of the vector meson contributions in the form factor $g(s)$:
\be
\alpha_{\omega K} \equiv \frac{f_{K^{'*}}g_{K^{'*}\omega 
K}}{f_{K^*}g_{K^*\omega K}},\ \ \ {\rm and}\ \ \ \alpha_{\phi K} \equiv 
\frac{f_{K^{'*}}g_{K^{'*}\phi K}}{f_{K^*}g_{K^*\phi K}} \ .
\ee
We can further attempt the use of SU(3) symmetry to derive such  
couplings. 
Instead, we will fix the values of $\alpha_{\omega K}$ by requiring that 
it reproduces the experimental branching fraction for the well measured 
$\tau^- \to \omega K^-\nu$ decay (see Table I). Using this method we 
obtain 
two possible values: $\alpha_{\omega K}= 0.54 \pm 0.38$ and 
$\alpha_{\omega  K}=-0.77 \pm 0.40$. Both are consistent with 
SU(3) since they have similar sizes to the value $\alpha_{\omega 
\pi}=-0.57 \pm 0.11$, which reproduces  the $\tau^- \to \omega \pi^-\nu$ 
decay data (see section 3).

   Now, if we assume that $\alpha_{\phi K} \approx \alpha_{\omega K}$ 
which is also expected on the basis of SU(3), we can predict:
\be
B(\tau^- \to \phi K^-\nu) = \left\{ \begin{array}{c} 
(2.2 \pm 2.6)\x  10^{-5}\ , \ \ \ {\rm for}\ \alpha_{\phi K}= 0.54 \pm 
0.38\ , \\
(1.6 \pm 2.5) \x 10^{-5}\ , \ \ \ {\rm for}\ \alpha_{\phi K}= -0.77 \pm
0.40\, \end{array} \right. 
\ee
which are consistent with the experimental values measured by BABAR 
\cite{Babar} and  BELLE \cite{Abe:2006uk} Collaborations (see Table I). 

Note that the large error bars quoted in Eq. (39) are dominated by the 
uncertainty in the $K'_1=K_1(1400)$ weak decay constant given in Eq. 
(13), which was extracted from the poorly measured $\tau \to 
K_1(1400)\nu$ decay. This  error bar can be reduced if we use a weak 
coupling constant obtained from  a phenomenological quark model. Thus, for 
example, if we assume $\alpha=45^0$, from the covariant quark model of 
ref. \cite{Cheng:2003sm} we obtain $f^{cqm}_{K'_1}=(242 \pm 25)\x 10^3$ 
MeV$^2$. This value of the weak coupling does not affect in a 
sensitive way the ratio of $K^{'*}/K^*$ couplings extracted from $\tau^- 
\to \omega K^-\nu$ branching fraction, which now become: 
$\alpha_{\omega K^-}=0.55 \pm 0.38$ or $\alpha_{\omega K^-}=-0.78 \pm 
0.39$. Using the value of $f_{K'_1}$ obtained above,  
we get: 
\be B^{cqm}(\tau^- \to \phi K^-\nu)= \left\{ \begin{array}{c}
(4.0 \pm 1.2)\x  10^{-5}\ , \ \ \ {\rm for}\ \alpha_{\phi K}= 0.55 \pm
0.38\ , \\
(3.3 \pm 1.0) \x 10^{-5}\ , \ \ \ {\rm for}\ \alpha_{\phi K}= -0.78 \pm
0.39\, \end{array} \right.
\ee
which central values are in better agreement with experimental data of 
BABAR and BELLE  (see Table I). 

\begin{figure}
\includegraphics[width=12cm]{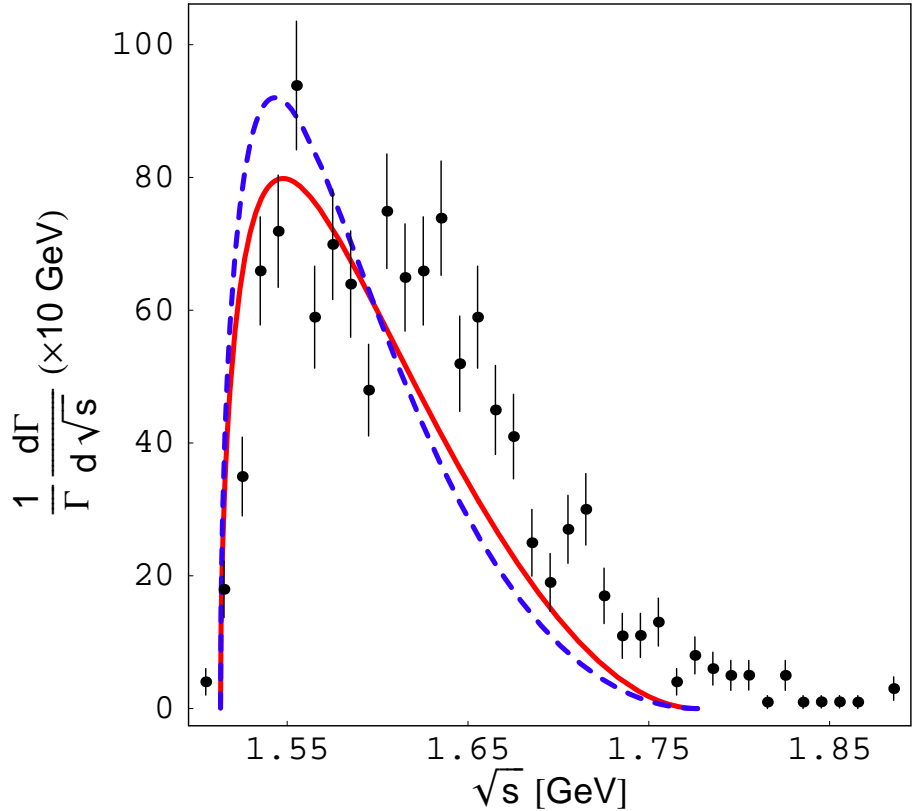}
%\vspace{-10.0cm}
\caption{Invariant mass distribution of the $\phi K^-$ system in tau 
decays. The solid (dashed) line corresponds to $\alpha_{\phi K}=0.55$ 
($\alpha_{\phi K}=-0.78$). Data points are taken from Ref. \cite{Babar}.} 
\end{figure}

   In Figure 3 we compare the invariant mass distribution of the $\phi 
K^-$ system, with the measurements reported by the BABAR Collaboration 
\cite{Babar}. As we can observe, our model (with our numbers multiplied 
by an arbitrary scale) nicely reproduces the data on the invariant mass 
distribution. Although it is difficult to discriminate between values 
of the two-fold ambiguity in $\alpha_{\phi K}$, data seems to favor the 
solution with $\alpha_{\phi K}=0.55$. 

\section{Conclusions}

  The decays $\tau^- \to (\omega,\ \phi)P^-\nu_{\tau}$, where $P$ 
is a charged pseudoscalar meson, are studied in a phenomenological model 
where the form factors are dominated by the exchange of meson states with 
the appropriate quantum numbers. We rely on SU(3) flavor symmetry to fix 
the strong interaction coupling constants, and assume an ideal value for 
the $\omega-\phi$ mixing angle. Our predictions for the decay modes 
involving a $\phi$ vector meson:
\ba
B(\tau^- \to \phi \pi^-\nu_{\tau})&=& (3.64 \pm 0.93) \x 10^{-5}, \nn \\
B(\tau^- \to \phi K^-\nu_{\tau})&=& (4.0 \pm 1.2) \x 10^{-5}, 
\ea
are in very good agreement with measurements reported recently by the 
BABAR \cite{Babar} and BELLE \cite{Abe:2006uk} Collaborations. 

\acknowledgments

The authors would like to thank the financial support from Conacyt 
(M\'exico).

\

\appendix{\large \bf Appendix}
	
There are two octets of axial mesons corresponding to the $1^3P_1$ 
(denoted by $A$)  and  $1^1P_1$ (denoted by $B$) quantum number 
configurations. The $A$ ($B$) octet is composed of one isotriplet $a_1$ 
($b_1$), two  isodoublets $K_{1A}$ ($K_{1B}$) and one isosinglet $f_1^8$ 
($h_1^8$) states. The axial-vector-pseudoscalar interaction is 
governed by the Lagrangian shown in Eq. (27), with physical strange mesons 
$K_1(1270)$ and $K_1(1400)$ defined in Eqs. (28)-(29).

   From the interaction Lagrangian (27) we can derive the following 
strong coupling constants that involve the $K_1,\ K'_1$ axial mesons and 
the $\omega,\ \phi$ vector mesons of our interest:
\begin{eqnarray*}
g_{K_1^+ \omega K^-} & = & g_{K_1^0 \omega \overline{K}^0} = - g_{K_1^-
\omega K^+} = - g_{\overline{K}_1^0 \omega K^0} \\
 & = &
\frac{\sqrt{3}}{2} \bigg[ g_{A V P}^8 \sin\alpha \sin\theta_V -
 \frac{1}{3} g_{B V P}^8 \cos\alpha \big( \sin\theta_V - 2 \sqrt{2} r_B
\cos\theta_V \big) \bigg] \ , \\
g_{K_1^{'+} \omega K^-} & = & g_{K_1^{'0} \omega \overline{K}^0} = -
g_{K_1^{'-} \omega K^+} = - g_{\overline{K}_1^{'0} \omega K^0} \\
 & = & - \frac{\sqrt{3}}{2} \bigg[ g_{A V P}^8 \cos\alpha \sin\theta_V +
 \frac{1}{3} g_{B V P}^8 \sin\alpha \big( \sin\theta_V - 2 \sqrt{2} r_B
\cos\theta_V \big) \bigg] \ , \\
 g_{K_1^+ \phi K^-} & = & g_{K_1^0 \phi \overline{K}^0} = - g_{K_1^- \phi
K^+} = - g_{\overline{K}_1^0 \phi K^0} \\
  & = & \frac{\sqrt{3}}{2} \bigg[ g_{A V P}^8 \sin\alpha \cos\theta_V -
 \frac{1}{3} g_{B V P}^8 \cos\alpha \big( \cos\theta_V + 2 \sqrt{2} r_B
\sin\theta_V \big) \bigg] \ , \\ 
 g_{K_1^{'+} \phi K^-} & = & g_{K_1^{'0} \phi \overline{K}^0} = -
 g_{K_1^{'-} \phi K^+} = - g_{\overline{K}_1^{'0} \phi K^0} \\
  & = & - \frac{\sqrt{3}}{2} \bigg[ g_{A V P}^8 \cos\alpha \cos\theta_V +
 \frac{1}{3} g_{B V P}^8 \sin\alpha \big( \cos\theta_V + 2 \sqrt{2} r_B
\sin\theta_V \big) \bigg] \ ,
\end{eqnarray*}
where we have defined the ratio of singlet and octet couplings $r_B 
\equiv g^0_{BVP}/g^8_{BVP}$. In the above expressions, $\theta_V$ 
(respectively $\alpha$) denotes  the $\omega-\phi$ ($K_1-K'_1$) mixing 
angle.

  Other useful couplings involving the $K_1$ and $K'_1$ axial mesons are:
\begin{eqnarray*}
g_{K_1^+ \rho^0 K^-} & = & g_{\overline{K}_1^0 \rho^0 K^0} = - g_{K_1^-
\rho^0 K^+} = - g_{K_1^0 \rho^0 \overline{K}^0} 
  =  \frac{1}{2} \big( g_{A V P}^8 \sin\alpha + g_{B V P}^8 \cos\alpha
\big) \ , \\
g_{K_1^+ \rho^- \overline{K}^0} & = & g_{K_1^0 \rho^+ K^-} = - g_{K_1^-
\rho^+ K^0} = - g_{\overline{K}_1^0 \rho^- K^+} 
  =   \frac{1}{\sqrt{2}} ( g_{A V P}^8 \sin\alpha + g_{B V P}^8
\cos\alpha ) \ , \\
g_{K_1^{'+} \rho^0 K^-} & = & g_{\overline{K}_1^{'0} \rho^0 K^0} = -
g_{K_1^{'-} \rho^0 K^+} = - g_{K_1^{'0} \rho^0 \overline{K}^0} 
  = - \frac{1}{2} ( g_{A V P}^8 \cos\alpha - g_{B V P}^8 \sin\alpha )
\ , \\
g_{K_1^{'+} \rho^- \overline{K}^0} & = & g_{K_1^{'0} \rho^+ K^-} = -
g_{K_1^{'-} \rho^+ K^0} = - g_{\overline{K}_1^{'0} \rho^- K^+} 
  =  - \frac{1}{\sqrt{2}} ( g_{A V P}^8 \cos\alpha - g_{B V P}^8
\sin\alpha ) \ , \\
g_{K_1^+ K^{*-} \pi^0} & = & g_{\overline{K}_1^0 K^{*0} \pi^0} = -
g_{K_1^- K^{*+} \pi^0} = - g_{K_1^0 \overline{K}^{*0} \pi^0}  
 =  - \frac{1}{2} ( g_{A V P}^8 \sin\alpha - g_{B V P}^8 \cos\alpha )
\ , \\
g_{K_1^+ \overline{K}^{*0} \pi^-} & = & g_{K_1^0 K^{*-} \pi^+} = -
g_{K_1^- K^{*0} \pi^+} = - g_{\overline{K}_1^0 K^{*+} \pi^-} 
  =  - \frac{1}{\sqrt{2}} ( g_{A V P}^8 \sin\alpha - g_{B V P}^8
\cos\alpha ) \ , \\
g_{K_1^{'+} K^{*-} \pi^0} & = & g_{\overline{K}_1^{'0} K^{*0} \pi^0} = -
g_{K_1^{'-} K^{*+} \pi^0} = - g_{K_1^{'0} \overline{K}^{*0} \pi^0} 
 =  \frac{1}{2} ( g_{A V P}^8 \cos\alpha + g_{B V P}^8 \sin\alpha ) \
, \\
g_{K_1^{'+} \overline{K}^{*0} \pi^-} & = & g_{K_1^{'0} K^{*-} \pi^+} = -
g_{K_1^{'-} K^{*0} \pi^+} = - g_{\overline{K}_1^{'0} K^{*+} \pi^-} 
   =  \frac{1}{\sqrt{2}} ( g_{A V P}^8 \cos\alpha + g_{B V P}^8
\sin\alpha ) \ .
\end{eqnarray*}

\end{document}